\documentclass[showpacs,twocolumn,amssymb,prl,aps]{revtex4}
\usepackage{graphicx,epsfig}

\begin{document}
\title{Tetra Point Wetting at the Free Surface of Liquid Ga-Bi}

\author{P.~Huber$^1\email{
Corresponding author. E-mail: patrick@xray.harvard.edu}$,
O.G.~Shpyrko$^1$, P.S.~Pershan$^1$, B.M.~Ocko$^2$, E.~DiMasi$^2$,
and M.~Deutsch$^3$}
\affiliation{$^1$Department of Physics, Harvard University, Cambridge MA 02138\\
$^2$Department of Physics, Brookhaven National Lab, Upton
NY 11973\\
$^3$Department of Physics, Bar-Ilan University, Ramat-Gan 52900,
Israel}

\date{\today}

\begin{abstract}
A continuous surface wetting transition, pinned to a
solid/liquid/liquid/vapor tetra coexistence point, is studied by
x-ray reflectivity in liquid Ga-Bi binary alloys. The short-range
surface potential is determined from the measured temperature
evolution of the wetting film. The thermal fluctuations are shown
to be insufficient to induce a noticeable breakdown of mean-field
behavior, expected in short-range-interacting systems due to their
$d_u=3$ upper critical dimensionality.
\end{abstract}


\maketitle

The wetting transition, predicted independently by
Cahn\cite{cahn1977} and by Ebner and Saam\cite{ebner1977} in 1977,
has attracted much theoretical\cite{DeGennes1985} and
experimental\cite{Law2001} attention, due to its importance for
fundamental physics\cite{DeGennes1985} and for
applications\cite{applic}. Almost all experimental and theoretical
studies of the wetting transition published to date address
liquids interacting through long-range van der Waals potentials,
for which the upper critical dimensionality is $d_{u}<3$. Thus,
for a $d=2$ surface on a $d=3$ bulk mean-field (MF) behavior is
expected. By contrast, wetting phenomena in liquids interacting by
a short range potential (SRW), where $d_{u}=3$, allow, in
principle, to explore the regime where the MF behavior breaks down
due to fluctuations and the renormalization-group (RG) approach
becomes applicable. However, due to extremely demanding
experimental requirements and the lack of a theory allowing to
predict the SRW interaction parameters only a single experimental
study addressing fluctuation effects on SRW was published to
date\cite{Ross1999}. Even that study employs
van-der-Waals-interacting liquids, and SRW was achieved by tuning
the wetting temperature, $T_{W}$, very close to the critical
point, $T_{C}$. The authors' estimates of the fluctuations'
magnitude predicted a RG behavior. Nevertheless, a MF behavior was
found and interpreted as an "apparent failure" of RG theory to
account for SRW.

We present here an x-ray study of wetting in a Ga-Bi binary alloy,
which interacts by a short range screened Coulomb potential at all
concentrations. This allows achieving SRW far from $T_C$, thus
minimizing any possible influence of the criticality on the
wetting transition. Moreover, the x-ray techniques employed
allowed the determination of the wetting layer's internal
structure on an {\AA}ngstr{\"o}m scale and demonstrated the strong
compositional gradient within the wetting film, predicted by
density functional calculations\cite{ebner1977,TelodaGama1983}.
The wetting transition in this system is pinned to a bulk
monotectic temperature, $T_{M}$, creating a rare case of tetra
point wetting, where four phase coexist in the bulk: a Ga-rich
liquid, a Bi-rich liquid, a solid Bi and a
vapor\cite{Dietrich1997,wynblatt}. We determined the temperature
variation of the wetting layer's composition, thickness and
internal structure along a trajectory in the phase diagram which
includes several phase boundaries and the tetra point. The study
highlights the intimate relation between the wetting layer's
structure and the bulk phase diagram. The T-dependence of the
wetting film's structure yields the first \textit{quantitative}
determination of a MF short-range surface potential governing the
wetting transition. Finally, from a RG analysis we estimate the
influence of thermal fluctuations on the observed tetra point
wetting.

\begin{figure}[!]
\epsfig{file=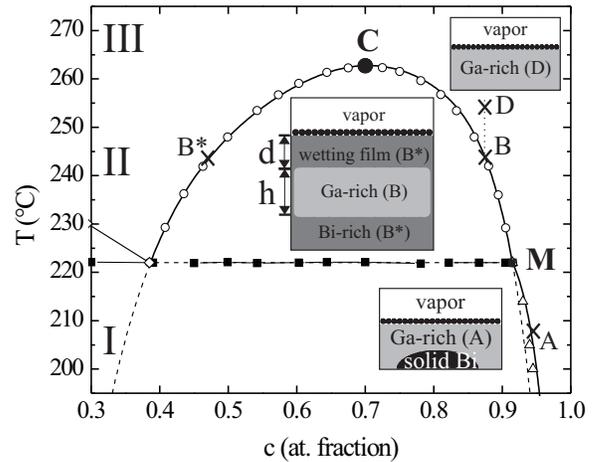, angle=0, width=1.00\columnwidth}
\caption{\label{fig1}The atomic fraction(c)-Temperature(T) bulk
phase diagram of Ga-Bi. The symbols indicate coexistence lines of
Predel's phase diagram\cite{predel1960}. The lines show the phase
boundaries calculated from thermodynamic data\cite{CALPHAD2000}.
The dashed lines represent the metastable extension of the l/l
coexistence line below T$_{M}$. The points are: C-bulk critical
point, M-monotectic point, A,B,D-points on the experimental path.
The insets illustrate the surface and bulk phases. In region II
the wetting film is 50 \AA\ thick and the Ga-rich fluid is 5 mm
thick. The bold surface lines in the insets symbolize the
Bi-monolayer.}
\end{figure}

The measured\cite{predel1960} bulk phase diagram of Ga-Bi is shown
in Fig.~\ref{fig1}, with cartoons relating the surface phases to
the bulk ones. For all $T$ a Gibbs-adsorbed Bi monolayer is found
at the free surface\cite{tostmann2000,Lei1996}. In Region I,
$T<T_{M}$, the Bi monolayer and the solid Bi coexist with a
Ga-rich liquid. In Region II, $T_{M}<T<T_{C}$, the bulk separates
macroscopically into immiscible Ga-rich and Bi-rich liquids, and
the denser Bi-rich phase sinks to the bottom of the sample pan.
However, the Bi-rich phase also wets the free surface by intruding
between the Ga-rich low density phase and the Bi monolayer, in
defiance of gravity. In the homogeneous part of the phase diagram,
region III, only the Bi monolayer is found at the free surface
\cite{Huber2001}. The four phases, solid Bi, Bi-rich liquid,
Ga-rich liquid, and vapor coexist at the boundary between regions
I and II, $T_{M}$, rendering it a solid-liquid-liquid-vapor tetra
point.

X-ray reflectivity measurements were carried out at beamline X22B,
NSLS, at a wavelength $\lambda =1.54\,$\AA . The intensity
$R(q_{z})$ reflected from the surface, is measured as a function
of the normal component $q_{z}$ of the momentum transfer and
allows determining the surface-normal electron density profile
$\rho (z)$\cite{Pershan1984}. The Ga-Bi alloy was prepared in an
inert-gas box using $>99.9999\%$ pure metals. A solid Bi pellet
was covered by an amount of liquid Ga required for a nominal
concentration $c_{nom}=88$ at\% Ga. It was then transferred in air
into an ultrahigh vacuum chamber. A day of bake-out yielded a
pressure of 10$^{-10}$ torr. The residual surface oxide on the
liquid's surface was removed by sputtering with Ar$^{+}$ ions.
Using thermocouple sensors and an active temperature control on
the sample pan and its adjacent thermal shield a temperature
stability and uniformity of $\pm 0.05^{\circ }C$ was achieved near
$T_{M}$.

\begin{figure}
\includegraphics{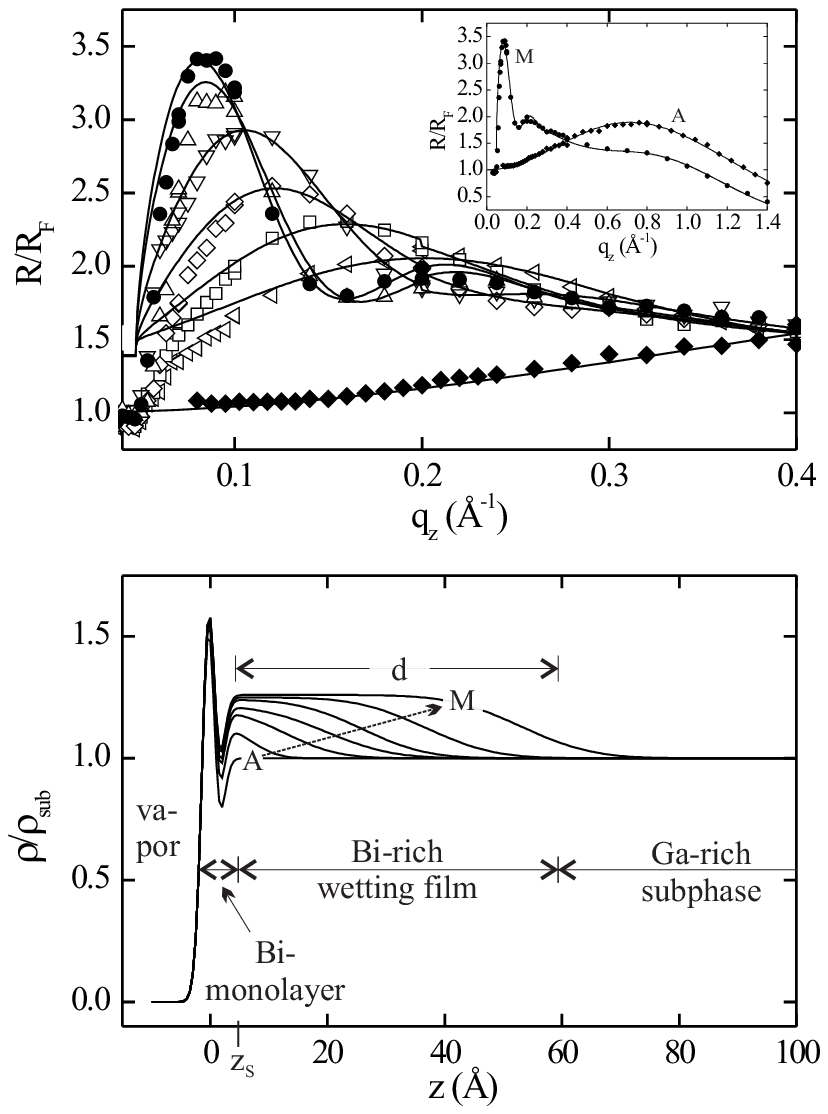}
\caption{\label{fig2} (a) $R/R_{F}$ for selected temperatures
$T_{A}\leq T\leq T_{M}$: $(\blacklozenge )$ $205.0^\circ C=T_{A}$,
$(\triangleleft)$ $218.9^\circ C$, $(\square)$ $220.4^\circ C$,
$(\lozenge)$ $221.0^\circ C$, $(\bigtriangledown)$ $221.5^\circ
C$, $(\bigtriangleup)$ $221.8^\circ C$, $(\bullet)$ $222.0^\circ C
=T_{M}$. Solid lines: fits to a two-box model of $\rho (z)$.
(b) Fit-refined electron density profiles $\rho (z)$ for aforementioned T's with $%
T_{A}\leq T\leq T_{M}$, $\rho _{sub}=$ electron density of Ga-rich
subphase. A and M mark $\rho (T_A)$ and $\rho (T_M)$,
respectively.}
\end{figure}

X-ray reflectivity $R(q_{z})$ was measured at selected
temperatures on the path A${\rightarrow}$M. The standard
procedure\cite{Pershan1994} for determining $\rho (z)$ from the
measured $R(q_{z})$ is to construct a physically motivated model
for $\rho (z)$ and fit its Fourier transform to the measured
$R(q_{z})/R_{F}(q_{z})$. $R_{F}(q_{z})$ is the Fresnel
reflectivity from an ideally flat and sharp surface having the
electron density of the Ga-rich liquid. We employ a two-box
model\cite{tostmann2000}, where the upper box represents the Gibbs
adsorbed Bi monolayer, and the lower box - the Bi-rich wetting
film. The model also includes three roughnesses for the three
interfaces: vapor/Bi-monolayer, Bi-monolayer/Bi-rich film, and
Bi-rich film/Ga-rich subphase. The fits (lines) to the measured
(points) $R/R_{F}$ are shown in Fig. \ref{fig2}(a), and the
corresponding $\rho (z)$ profiles - in Fig.~\ref{fig2}(b). At
point A ($T_{A}=205^{\circ }$C, inset to Fig.\ref {fig2}(a)),
typical of region I, $R/R_{F}$ exhibits a pronounced maximum at
$q_{z}=0.8\AA ^{-1}$, indicative of a high electron density layer
at the surface. The $\rho (z)$ obtained from the fit is consistent
with a segregated monolayer of pure Bi at the free
surface\cite{tostmann2000,Lei1996}. As the temperature is
increased towards M, a peak in $R/R_{F}$ starts to develop at
$q_{z}\approx 0.2${\AA}$^{-1}$, gradually shifting to lower
$q_{z}$. This behavior manifests the formation of the wetting
layer and its continuous growth in thickness upon approaching
$T_{M}$. For the fully-formed wetting layer at M
($T_{M}=222.0^{\circ }$C), $R/R_{F}$ shows two low-$q_{z}$ peaks,
characteristic of a thick film, and the fit yields a maximal film
density of $\rho /\rho _{subphase}=1.26$ as shown in Fig.
\ref{fig2}(b). This agrees well with the value of $1.25$
calculated for this ratio from the phase diagram at $T_{M}$. The
fit-refined $\sim 50\AA$ film thickness also agrees with
ellipsometric measurements\cite{nattland1995}. As can be seen in
Fig.~\ref{fig2}(b), the evolution of the density profiles along
the path A$\rightarrow$M proceeds via highly inhomogeneous wetting
films that are similar to the microscopic density profiles
calculated using density functional
theory\cite{ebner1977,TelodaGama1983}.

\begin{figure}
\includegraphics{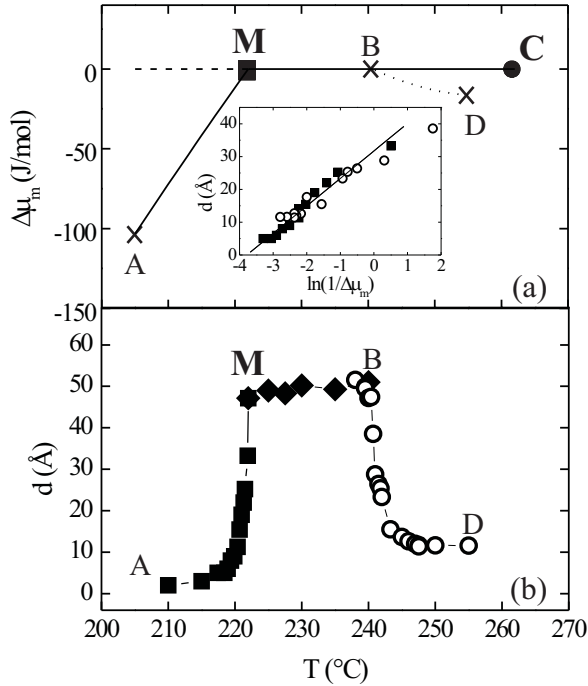}
\caption{\label{fig3} (a) ($\Delta \mu_{m}$,T) phase diagram:
(A-M) is the liquid-solid, and (M-C) is the liquid/liquid
coexistence lines. The path B-D is in the single phase region, and
M and C are the monotectic and critical points. Inset: effective
wetting layer thickness d on A$\rightarrow $M (squares) and
B$\rightarrow $D (open circles). The solid line is a fit to the
A$\rightarrow $M d values. (b) The measured d along the
experimental path.}
\end{figure}

The surface and bulk transitions at $T_{M}$ are driven by the
excess free energy, $\Delta \mu_{m}$, of the metastable Bi-rich
phase over that of the Ga-rich phase. Thus, to relate the observed
continuous surface transition to the first-order bulk transition
at $T_{M}$, we replot in Fig. \ref{fig3}(a) the $(c,T)$ phase
diagram of Fig.~\ref{fig1} on the $(\Delta \mu_{m},T)$ plane. For
$T>T_{M}$, the l/l coexistence line transforms into a straight
line at $(\Delta \mu_{m}=0$,T) extending from M to C. For
$T<T_{M}$ the metastable l/l coexistence line (dashed line)
extends horizontally below $T_{M}$, while the solid-Bi/Ga-rich
coexistence line goes below the metastable line from M to A. This
illustrates Dietrich and Schick's observation\cite{Dietrich1997}
that the path leading to coexistence on heating from A to M is
dictated by the topology of the phase diagram and corresponds to a
path which probes ''complete wetting'': $\Delta\mu_{m} \rightarrow
0$ on the path A $\rightarrow$ M.

To determine the surface potential, we define an effective wetting
film thickness,
$d=\int_{z_{s}}^{\infty}[{\rho}(z)-{\rho}_{Ga-rich}]/[{\rho
}_{Bi-rich}-{\rho }_{Ga-rich}]dz$. Here $z_{s}$ is the top of the
wetting film as marked in Fig.~\ref{fig2}(b), and ${\rho
}_{Bi-rich}$, ${\rho }_{Ga-rich}$ are the electron densities of
the coexisting bulk liquid phases, calculated from the phase
diagram. The $d$ values calculated from the reflectivity fits
along A$\rightarrow $M, M$\rightarrow $B, and
B$\rightarrow $D are plotted in Fig.~\ref{fig3}(b). Along A$%
\rightarrow $M a continuous increase in $d$ is observed, while on
the on-coexistence M$\rightarrow $B path a constant $d\approx 50$
\AA\ is found. This is in agreement with the predictions for SRW:
a continuous $d$ growth limited by gravity effects\cite{Law2001}.
At B the Ga concentration reaches its nominal value, $c_{nom}=88$
at\%, in the homogeneous liquid phase (region III). Upon further
increase of $T$ $c$ remains unchanged. The sample leaves the l/l
coexistence line, moving further into region III, and the wetting
film vanishes continuously. Further details of the M$\rightarrow
$D path are discussed elsewhere\cite{Huber2001}.

As implied by the similarity to Ebner and Saam's\cite{ebner1977}
results, a detailed calculation of the wetting film's
inhomogeneity will require either a density functional analysis,
or some other equivalent approach. Nevertheless, even a simple
model approximating the wetting layer by a slab of thickness $d$
allows a confident determination of the surface potential. For
this model, the grand canonical potential $\Omega _{S}$ at the
surface is given in the MF approximation by $\Omega
_{S}=N\,A_{0}\,[d\,\Delta\mu+\xi_{MF}\,\Phi\, e^{-d/\xi_{MF} }]$
with a short-range potential of decay length $\xi_{MF} $ and
amplitude $\Phi $. Here $\Delta \mu=\Delta \mu _{m}+ \Delta \mu
_{g}$ is the excess Gibbs Free Energy of the Bi-rich wetting phase
over that of the Ga-rich phase, and $\Delta \mu_{g}=\, g\, \Delta
\rho \,\,h\,\,V_{m}$ is the gravitational energy paid for having
the heavier Bi-rich phase at the surface. The particle number
density per unit volume is $N$,  $A_{0}$ is an arbitrary surface
area, $\,\Delta \rho $ is the mass density difference of the two
phases, $ V_{m}$ is the molar volume of the Bi-rich phase, and $h$
is the height difference between top and bottom. The equilibrium
thickness, $d^{\ast }$, is that minimizing $\Omega _{S}$:
$\frac{\partial \Omega _{S}}{\partial d}|_{d^{\ast
}}=0\,$\cite{DeGennes1985}. Neglecting the gravitational term for
points off l/l coexistence yields: $d^{\ast }=\xi_{MF} \,\ln
\left( \Phi /\Delta \mu_{m}\right) $. The inset in
Fig.~\ref{fig3}(a) shows the experimental $d$ values on the path
A$\rightarrow$M (solid squares). The clear linear dependence on
$\ln (1/\Delta \mu )$ confirms the major theoretical predictions
for SRW: a continuous growth and a logarithmic divergence of the
wetting film's thickness\cite{DeGennes1985}. A fit of $d^{\ast }$
(line) to the experimental $d$ (solid points) yields
$\xi_{MF}=6.3$ \AA\ and $\Phi =43$ J/mol. Both values are
reasonable for such a metallic system. The value of $\Phi$
corresponds to a change in surface energy upon wetting of about
$400$ mJm$^{-2}$ which is consistent with surface energy
measurements on Ga-Bi alloys\cite{Khokonov}, and Ga-Pb ones
\cite{wynblatt}. It is interesting to compare the behavior of $d$
for the path A$\rightarrow$M with that along the path
B$\rightarrow$D, shown in the inset of Fig.~\ref{fig3} (open
circles). The values overlap for most of the path, demonstrating
that the behavior along A$\rightarrow$M is the same as can be
observed for any other path probing complete wetting, e.g. the
path B$\rightarrow$D which is solely determined by the choice of
$c_{nom}$.

The analysis above exhibits a very good agreement of the
experimental SRW results with the universal behavior predicted by
MF theory. However, it is important to note that this does not
automatically mean a failure of the RG predictions for a
non-universal SRW behavior in our, and other, $d_{u} =3$ systems.
Both MF and RG approaches yield here a continuous, logarithmic
divergence of $d^{\ast }$ at $T_{W}$, but with different
amplitudes for $d^{\ast}$\cite{DeGennes1985,RGA}. For MF, the
amplitude is $\xi_{MF}$ while for RG, where $d^{\ast
}\sim\xi_{RG}\,(1+\omega /2)\, \ln\, (1 /\Delta \mu)$, the
amplitude is $\xi_{RG}\,(1+\omega /2)$. The "capillary parameter"
$\omega = k_{B} T_{W}/(4 \pi \xi_{b}^2 \gamma)$ (where $\xi_b$ is
the bulk correlation length, and $\gamma$ is the l/l interfacial
tension) measures the magnitude of the dominant thermal
fluctuations in the system, the thermally induced capillary waves
at the l/l interface of the coexisting Bi- and Ga-rich liquids.
Hence, $\omega$ measures the deviation from a MF behavior, which,
of course, neglects all fluctuations. A calculated\cite{liqliq}
$\gamma=4mJm^{-2}$ and a $\xi_b =6.1 \pm 1$ \AA\  estimated from
the two-scale factor universality (TSFU) theory for bulk demixing
in Ga-Bi\cite{Kreuser1993}, yield $\omega(T_{M})=0.3\pm0.2$. This
is well within the $\omega < 2$ range, where RG analysis predicts
only a mild effect on complete wetting\cite{RGA}. Indeed,
$\xi_{RG}=5.4 $ {\AA} is only $\sim 15$ {\%} smaller than
$\xi_{MF}=6.3$ \AA~above. The two values are close, and in
reasonable agreement with the value of $\xi_b$ expected from the
TSFU\cite{DeGennes1985}. Thus, a clear distinction between RG and
MF behavior can not be drawn in our case.

Finally, we compare the measured gravity-limited maximal thickness
of 50 \AA\ along M$ \rightarrow $B with theory. On coexistence
$\Delta \mu _{m}=0$ and thus $d_{g}^{\ast }=\xi_{RG} \,\ln \left(
\Phi /\Delta \mu_{g}\right)$. Using $\Phi =43$ J/mol and the known
material constants in $ \Delta \mu_{g}$ yields
$d_{grav}^{\ast}=15.6\,\,\xi_{RG}\, \approx 85$ \AA. Since this
analysis does not take into account the excess energy associated
with concentration gradients across the interfaces some
overestimation of $d_g^{\ast}$ is not too surprising. However,
even this rough calculation shows that the wetting film thickness
is expected to be on a mesoscopic rather than on the macroscopic
length scale observed for similar wetting geometries in systems
governed by long-range, dispersion forces\cite{Law2001}.

We presented here the first detailed, {\AA}ngstr{\"o}m resolution
study of a short-range wetting transition at the free surface of
the binary Ga-Bi liquid alloy. The characteristic properties of
SRW, a logarithmic divergence of the wetting layer's thickness
upon approaching the transition, are clearly exhibited by the
measurements. The observed strongly inhomogeneous internal
structure of the wetting layer is similar to that obtained from
density functional calculations of wetting at a hard
wall\cite{ebner1977,TelodaGama1983}. The surface wetting
transition is found to be pinned to the bulk monotectic point M in
this alloy, and the topology of the bulk phase diagram enforces a
path probing complete wetting\cite{Dietrich1997}. Since this path
ends in M where four phase coexist, the surface transition
constitutes a rare tetra point wetting effect. An estimation of
the capillary parameter $\omega$ indicates that fluctuations
affect the observed \textit{complete} SRW transition only
marginally. A \textit{critical} wetting transition with a similar
value of $\omega$ should show more pronounced deviations from MF
behavior. Here, however, this critical transition occurs on the
metastable extension of the l/l coexistence line somewhere below
$T_{M}$, and may be reachable only if the alloy supercools
sufficiently. Due to the mild effect of the fluctuations, the
surface potential driving the transition could be determined
quantitatively. This is of particular importance since in contrast
with van-der-Waals interacting liquids, where the surface
potential can be estimated from the
Dzyaloshinski-Lifshitz-Pitaevskii theory\cite{Law2001,Dzyalo1961},
no equivalent theory is available for liquids with short-range
interaction. We hope that this first quantitative determination of
the SRW potential will stimulate the development of such theories.
Available predictions, in turn, would initiate more experimental
studies of fluctuation effects in SRW. In particular, they may
allow a quantitative investigation of Parry's prediction of an
additional renormalization of $\omega$ in SRW due to a coupling
between the fluctuations at the liquid/vapor and at the l/l
interfaces\cite{Parry1996}. This coupling may resolve the
dichotomy between RG prediction and the observations in the single
previous experimental study of SRW\cite{Ross1999}.

\acknowledgements We thank Prof. S. Dietrich and Prof. B.I.
Halperin for helpful discussions. This work is supported by U.S.
DOE Grant No. DE-FG02-88-ER45379, NSF Grant NSF-DMR-98-72817, and
the U.S.-Israel Binational Science Foundation, Jerusalem. BNL is
supported by U.S. DOE Contract No. DE-AC02-98CH10886. PH
acknowledges support from the Deutsche Forschungsgemeinschaft.



\begin{references}
\bibitem{CALPHAD2000}  The phase boundaries were calculated using data from the CalPhaD initiative (L.
Kaufmann, MIT).
\end{references}
\end{document}